\documentclass[11pt]{article}
  
\usepackage{amsmath,amssymb}

\usepackage{epsfig}  
\usepackage{graphicx}               % Standard graphics package  
\usepackage{url}
\usepackage{hyperref}
\usepackage{cite}

\usepackage{color}

\newcommand{\nc}{\newcommand}  

\nc{\beq}{\begin{equation}}  
\nc{\eeq}{\end{equation}}  
\nc{\beqa}{\begin{eqnarray}}  
\nc{\eeqa}{\end{eqnarray}}  
\nc{\bea}{\begin{eqnarray}}  
\nc{\eea}{\end{eqnarray}}  
\nc{\ra}{\rightarrow}  
\nc{\lsim}{\begin{array}{c}\,\sim\vspace{-21pt}\\< \end{array}}  
\nc{\gsim}{\begin{array}{c}\sim\vspace{-21pt}\\> \end{array}}  
\nc{\Tr}{{\rm Tr}}
\nc{\slsh}{\slash\hspace*{-0.22cm}}

\def\be{\begin{equation}}
\def\ee{\end{equation}}
\def\bea{\begin{eqnarray}}
\def\eea{\end{eqnarray}}
\def\bit{\begin{itemize}}
\def\eit{\end{itemize}}

\def\to{\rightarrow}

\title{  
\vspace*{-2.3cm}  
\begin{flushright}  
\normalsize{  
ANL-HEP-PR-13-30 
  }  
\end{flushright}  
\vspace{1.5cm}  
\Large  
\textbf{
The Scale of Dark QCD
}\vspace*{1.0cm}   
}

\author{Yang Bai~$^{a}$ and Pedro Schwaller~$^{b,c}$
\vspace{5mm}
\\
$^{a}$ \normalsize\emph{Department of Physics, University of Wisconsin, Madison, WI 53706, USA}  \vspace{1mm} \\
$^{b}$ \normalsize\emph{HEP Division, Argonne National Laboratory, 9700 Cass Ave., Argonne, IL 60439, USA}
\vspace{1mm} \\
$^{c}$ \normalsize\emph{Physics Department, University of Illinois at Chicago, Chicago, IL 60607, USA}
}

\date{}

\begin{document}  

\setcounter{page}{0}  
\maketitle  

\vspace*{1cm}  
\begin{abstract} 
Most of the mass of ordinary matter has its origin from quantum chromodynamics (QCD). A similar strong dynamics, dark QCD, could exist to explain the mass origin of dark matter. Using infrared fixed points of the two gauge couplings, we provide a dynamical mechanism that relates the dark QCD confinement scale to our QCD scale, and hence provides an explanation for comparable dark baryon and proton masses. Together with a mechanism that generates equal amounts of dark baryon and ordinary baryon asymmetries in the early Universe, the similarity of dark matter and ordinary matter energy densities can be naturally explained. For a large class of gauge group representations, the particles charged under both QCD and dark QCD, necessary ingredients for generating the infrared fixed points, are found to have masses at 1-2~TeV, which sets the scale for dark matter direct detection and novel collider signatures involving visible and dark jets. 
\end{abstract}  
  
\thispagestyle{empty}  
\newpage  
  
\setcounter{page}{1}

\baselineskip18pt   

\vspace{-3cm}

%%%%%%%%%%%%%%%%%%%%%%%%%%%%%%%%%%%%
\section{Introduction}
\label{sec:intro}
%%%%%%%%%%%%%%%%%%%%%%%%%%%%%%%%%%%%
Over the last few decades, cosmological observations have firmly established that an unknown form of matter, \textit{dark matter} (DM), is present in the Universe. Within the context of the standard $\Lambda$CDM cosmology, the recent Planck data has determined the cold dark matter energy density to the highest precision: $\Omega_{\rm DM} h^2 = 0.1199\pm0.0027$~\cite{Ade:2013zuv}, which is a factor of $\Omega_{\rm DM}/\Omega_{\rm B}=5.44\pm0.14$ times the baryon energy density. To explain the cold dark matter energy density, weakly interacting massive particles have served as the leading candidate~\cite{Feng:2010gw}. Their masses are related to the electroweak scale and their number density or relic density is from a thermal freeze-out mechanism. 

Ordinary matter, on the other hand, has its mass coming from the proton (or neutron) mass $m_p$, which is related to the QCD confinement scale $\Lambda_{\rm QCD}$ in the Standard Model (SM). Its number density 
originates from a baryon-antibaryon asymmetry.
Because of the similarity of the dark matter and ordinary matter energy densities, it is very likely that a strong dynamics similar to QCD exists in the dark matter sector and the dark matter energy density follows the same story as in our QCD sector. The dark matter energy density would then be a product of the dark baryon mass $m_{D}$ and its number density $n_{D}$. To have  comparable $\Omega_{\rm DM}$ and $\Omega_{\rm B}$, one needs to have $m_D \sim m_p$ if a common asymmetry mechanism provides $n_D \sim n_B$, which can be realized through many mechanisms~\cite{Nussinov:1985xr,Kaplan:1991ah,Barr:1990ca,Barr:1991qn,Dodelson:1991iv,Fujii:2002aj,Kitano:2004sv,Farrar:2005zd,Gudnason:2006ug, Kitano:2008tk,Kaplan:2009ag,Shelton:2010ta, Davoudiasl:2010am, Buckley:2010ui,Blennow:2010qp,Cohen:2010kn,Frandsen:2011kt} (see~\cite{Petraki:2013wwa} for a recent review). It is less trivial to have the dark matter mass comparable to the proton mass or the QCD scale. In this paper, we are trying to provide a natural explanation of $m_D \sim m_p$, or equivalently, for having the dark QCD scale comparable to our QCD scale, $\Lambda_{\rm dQCD}\sim \Lambda_{\rm QCD}$.

Given a new QCD-like dynamics in the dark sector, the dark QCD confinement scale $\Lambda_{\rm dQCD}$ depends on both the gauge coupling value at the far UV and its beta function from the matter content. Even if one chooses the same dark QCD gauge coupling as our QCD coupling in the UV, the exponential dependence of the confinement scale on the beta function can still generate $\Lambda_{\rm dQCD}$ far away from $\Lambda_{\rm QCD}$. Unless the dark QCD sector is an exact copy of our QCD sector, which would be a big surprise, additional mechanisms are required to have the dark QCD and our QCD couplings related to each other at a scale not too high, in order to suppress the renormalization running effects. 

\begin{figure}[th!]
\begin{center}
%\hspace*{-0.75cm}
\includegraphics[width=0.7\textwidth]{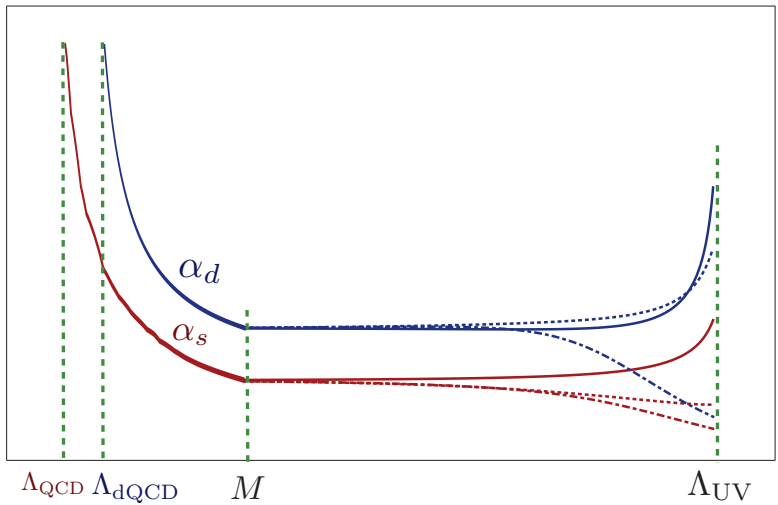}
\caption{An illustrative picture of the gauge coupling runnings from a UV scale to the confinement scales. Different UV boundary gauge couplings can lead to the same IRFPs. After decoupling particles charged under both groups at a scale $M$, both couplings run again below $M$ and generate compatible confinement scales $\Lambda_{\rm QCD}$ and $\Lambda_{\rm dQCD}$. 
}
\label{fig:running}
\end{center}
\end{figure}

For a single non-Abelian gauge group, increasing the multiplicity of matter content can suppress the first term in the beta function and potentially generate a nontrivial and perturbative infrared fixed point (IRFP)~\cite{Banks:1981nn}. For the QCD and dark QCD gauge groups, $SU(N_c)_{\rm QCD}\times SU(N_d)_{\rm dQCD}$, the two gauge couplings can have coupled beta functions as well as related IRFP values, $\alpha_s^*$ and $\alpha_d^*$, if some matter fields are charged under both gauge groups. Since we have not observed any additional particles charged under QCD below around the top quark mass, the fields charged under both gauge groups should have a mass at a scale $M \gtrsim m_t$ (for simplicity, we assume a common mass for them). Below the scale $M$, the IRFPs will be lifted and both gauge couplings run independently to generate $\Lambda_{\rm QCD}$ and $\Lambda_{\rm dQCD}$, respectively. The requirement of $\Lambda_{\rm dQCD}\sim \Lambda_{\rm QCD}$ from the dark matter energy density will prefer to have $M$ not too far away from $\Lambda_{\rm QCD}$ or $\Lambda_{\rm dQCD}$. To illustrate our idea, we show a schematic representation of gauge coupling runnings from a far UV scale, for instance the Planck scale, to the confinement scales in Fig.~\ref{fig:running}.
Once the field content is fixed, the gauge couplings at the IRFPs are also fixed. Our QCD gauge coupling value at $M_Z$ and the IRFP gauge coupling $\alpha_s^\star$ at $M$ can then be used to determine the decoupling scale $M$. Once $M$ is determined, the dark QCD running below $M$ is also known and the dark QCD confinement scale can be determined. The dark QCD scale, $\Lambda_{\rm dQCD}$, is therefore related to our QCD scale and fully determined by the field content. 

Our paper is organized as follows. In Section~\ref{sec:scale}, we study many models that can provide IRFPs and relate $\Lambda_{\rm dQCD}$ to $\Lambda_{\rm QCD}$. We pay special attention to the distribution of the scale $M$ for different models that satisfy $\Lambda_{\rm dQCD}\sim \Lambda_{\rm QCD}$.  In Section~\ref{sec:leptogenesis}, we construct a concrete renormalizable model to relate the dark baryon  to the ordinary baryon number densities and to explain the ratio $\Omega_{\rm DM}/\Omega_{\rm B}$. We discuss collider signatures and the dark matter direct detection rate of this model in Section~\ref{sec:pheno} and conclude in Section~\ref{sec:conclusion}.

%%%%%%%%%%%%%%%%%%%%%%%%%%%%%%%%%%%%
\section{The Scale of Dark QCD}
\label{sec:scale}
%%%%%%%%%%%%%%%%%%%%%%%%%%%%%%%%%%%%
Assuming an asymptotic-free QCD-like dynamics in the dark sector, the dark baryon in this sector could be a stable particle and serve as a dark matter candidate. Neglecting the electroweak symmetry, we have the gauge group $SU(N_c)_{\rm QCD} \times SU(N_d)_{\rm dQCD}$.
%
%\beqa
%G_{\rm gauge} = SU(N_c)_{\rm QCD} \times SU(N_d)_{\rm dQCD} \,.
%\eeqa
%
For simplicity, we only consider the case $N_d=N_c=3$ and fundamental representations for fermions and scalars under the gauge group. Other representations will not change the generic conclusions of this paper. Other than $n_{f_c}~(n_{f_d})$ Dirac fermions and  $n_{s_c}~(n_{s_d})$ complex scalars as fundamentals of $SU(N_c)~[SU(N_d)]$, we also introduce $n_{f_j}$ Dirac fermions and $n_{s_j}$ complex scalars as bifundamentals of $SU(N_c)\times SU(N_d)$, which are crucial to relating the IRFP gauge couplings in the two sectors. The particle content is summarized in Tab.~\ref{tab:matter-content}.
\begin{table}[!tb]
 \centering
  \renewcommand{\arraystretch}{1.3}
  \begin{tabular}{cccc}
    \hline \hline
    Field & $SU(N_c)_{\rm QCD}$ & $SU(N_d)_{\rm dark QCD}$  & Multiplicity  \\
    \hline \hline
   SM fermion  &   $N_c$    & 1     &  $n_{f_c}$  \\ \hline
   SM scalar    &    $N_c$   & 1     &  $n_{s_c}$ \\ \hline \hline
   DM fermion  &    1   & $N_d$     &  $n_{f_d}$ \\ \hline   
   DM scalar     &    1  & $N_d$     &  $n_{s_d}$ \\ \hline \hline    
   Joint fermion  &    $N_c$   & $N_d$     &  $n_{f_j}$ \\ \hline   
   Joint scalar     &    $N_c$  & $N_d$     &  $n_{s_j}$ \\ \hline    
    \hline
      \end{tabular}
  \caption{Matter content of the model. Multiplicities are for Dirac (vector-like) fermions and complex scalars. In particular, $n_{f_c} \geq 6$ to accommodate the SM quarks. Fields that are neutral under $SU(N_c)_{\rm QCD} \times SU(N_d)_{\rm dQCD}$ are not shown. 
  }
  \label{tab:matter-content}
\end{table}

At two-loop level, the two gauge couplings $g_c$ and $g_d$ affect each other's running. Defining the beta functions as $d g_c/d (\log{\mu}) = \beta_c(g_c, g_d)$ and $d g_d/d (\log{\mu}) = \beta_d(g_c, g_d)$, we have the beta functions at the two-loop level as~\cite{Jones:1981we}~\footnote{In Ref.~\cite{Jones:1981we}, chiral fermions are used. In our notation, we use Dirac fermions, so there is an additional factor of two in the formula.}
\beqa
\beta_c(g_c, g_d) &=& \frac{g_c^3}{16\pi^2}\,
\left[ \frac{2}{3}\,T(R_f)\,2(n_{f_c} + N_d\,n_{f_j} ) + \frac{1}{3}\,T(R_s)\, (n_{s_c} + N_d\,n_{s_j}) - \frac{11}{3}\,C_2(G_c) \right]
\nonumber \\
&&+\,\frac{g_c^5}{(16\pi^2)^2}\,\left[ 
\left(\frac{10}{3} C_2(G_c) + 2 C_2(R_f) \right)\,T(R_f)\,2\,(n_{f_c} + N_d \,n_{f_j} )  \right. \nonumber \\
&&\left. \qquad\qquad\quad \,+\, \left(\frac{2}{3}C_2(G_c) + 4 C_2(R_s) \right)\,T(R_s)\,(n_{s_c} + N_d\,n_{s_j})\,-\, \frac{34}{3}\,C^2_2(G_c) \right]     \nonumber \\
&&+\,\frac{g_c^3\,g_d^2}{(16\pi^2)^2} \left[
2 C_2(R_f)\,T(R_f)\,2\,N_d\,n_{f_j}\,+\,4 C_2(R_s)\,T(R_s)\,N_d\,n_{s_j}
\right] \,.
\eeqa
The formula for $\beta_d(g_c, g_d)$ is obtained from $\beta_c(g_c, g_d)$ by interchanging the indexes $c\leftrightarrow d$. Here, $C_2(G_c) = N_c$ and $C_2(G_d) =N_d$ are the quadratic Casimirs of the adjoint representations; $C_2(R_f)=C_2(R_s)=(N_{c, d}^2-1)/(2 N_{c, d})$ are  the quadratic Casimirs of  the fundamental representations; $T(R_f)=1/2$ and $T(R_s)=1/2$. We have checked and found that the electroweak gauge couplings and the top Yukawa coupling have negligible effects on the QCD and dark QCD couplings in the infrared. Similarly to the Banks-Zaks fixed point for a single gauge coupling~\cite{Banks:1981nn}, one can solve the zero beta-function equations $\beta_{c,d}(g_c,g_d) =0$ and obtain the perturbative IRFP as
\beqa
\alpha_s^* &\equiv& \alpha_s^* ( n_{f_c}, n_{s_c}, n_{f_d}, n_{s_d}, n_{f_j}, n_{s_j} ) \, , \nonumber \\
\alpha_d^* &\equiv& \alpha_d^* ( n_{f_c}, n_{s_c}, n_{f_d}, n_{s_d}, n_{f_j}, n_{s_j} ) \,,
\eeqa
with $\alpha_s= g_c^2/4\pi$ and $\alpha_d=g_d^2/4\pi$. Here, we assume that there are no masses for the fermions and scalars  between the UV scale and a lower scale of $M$ and no threshold corrections for the IRFP calculation. Assuming a common mass $M$ for all scalars and fermions except the QCD quarks and dark fermions charged only under dark QCD, the QCD coupling values, $\alpha_s(M)=\alpha_s^*$ and $\alpha_s(M_Z)= 0.1197\pm 0.0016$~\cite{Beringer:1900zz,Dissertori:2009ik}, can be used to determine the decoupling scale $M$. For some representative models, we show the IRFP gauge coupling values and the decoupling scale $M$ in Table~\ref{tab:fixed-point-values}.
\begin{table}[!tb]
 \centering
  \renewcommand{\arraystretch}{1.3}
  \begin{tabular}{c|cccccc|cc|c|c}
    \hline \hline
   Model & $n_{f_c}$ & $n_{f_d}$ & $n_{f_j}$  &    $n_{s_c}$ & $n_{s_d}$ & $n_{s_j}$  & $\alpha^*_s$ &  $\alpha^*_d$ & $M$~(GeV) & $m_D$~(GeV) \\
    \hline \hline
    A & 6 & 5 & 3 & 0 & 2 & 0 & 0.095 & 0.175 & 518 & 31 \\ \hline
  B & 6 & 6 & 3 & 1 & 0 & 0 & 0.083 & 0.120 & 2030  & 8.6  \\ \hline
 % C & 6 & 6 & 3 & 1 & 1 & 0 & 0.091 & 0.091 & 820 & 0.32 \\ \hline
%  D & 6 & 6 & 3 & 1 & 2 & 0 & 0.098 & 0.063 & 405 & $\approx 0$ \\ \hline
  C & 6 & 6 & 3 & 2 & 2 & 0 & 0.070 & 0.070 & 13500 & 0.32 \\ \hline
%  F & 6 & 7 & 3 & 2 & 0 & 0 & 0.082 & 0.022 &  2244   & $\approx 0$    \\ \hline
  D & 7 & 7 & 2 & 2 & 0 & 2 & 0.078 & 0.168 & 3860 & 72 \\ \hline
  E & 7 & 7 & 2 & 2 & 1 & 2 & 0.090 & 0.133  & 869 &  3.5  \\ \hline
%  I & 8 & 8 & 2 & 0 & 1 & 2 & 0.073 & 0.039 & 8050& $\approx 0$  \\ \hline
%  J & 8 & 8 & 2 & 0 & 2 & 2 & 0.081 & 0.015 & 2540 & $\approx 0$ \\ \hline
  F &  8 & 8 & 2 & 2 & 0 & 1 & 0.074 & 0.149 & 7700  &  29  \\ \hline
  G & 8 & 8 & 2 & 2 & 1 & 1 & 0.082 & 0.118 &  2244   & 1.2     \\ \hline
 % I & 8 & 8 & 2 & 2 & 2 & 1 & 0.089 & 0.089 &   970  &   0.02   \\ \hline
    \hline
      \end{tabular}
  \caption{The perturbative IRFP coupling values, decoupling scale $M$ and the dark baryon mass $m_D$ for some representative models. Matter fields that are charged under both gauge symmetries decouple at a mass scale $M$, which is determined from $\alpha_s^*$ and $\alpha_s(M_Z)= 0.1197\pm 0.0016$~\cite{Beringer:1900zz,Dissertori:2009ik}.}
  \label{tab:fixed-point-values}
\end{table}

Once the scale $M$ and the dark QCD coupling value $\alpha_d(M)=\alpha_d^*$ are known, we calculate the dark QCD gauge coupling from the scale $M$ to a lower scale. Because the gauge coupling $\alpha_d$ increases as the scale decreases, at a lower scale the dark QCD coupling can be large enough to trigger confinement and chiral symmetry breaking. The actual determination of such a scale requires a nonperturbative calculation. As a guidance, we use the chiral symmetry breaking condition from Cornwall, Jackiw and Tomboulis effective potential~\cite{Cornwall:1974vz}, which has $\alpha_d\, C_2(R_f) > \pi/3$ or $\alpha_d > \pi/4$~\cite{PeskinChiral}. From this condition, we define the dark QCD scale through the relation $\alpha_d (\Lambda_{\rm dQCD}) =\pi/4$. Applying the same calculation to our QCD scale, we have the relation between the proton mass and $\Lambda_{\rm QCD}$ as $m_p \approx 1.5 \,\Lambda_{\rm QCD}$. We apply this relation to the dark QCD and obtain the dark matter (dark baryon) mass as $m_D \approx 1.5\, \Lambda_{\rm dQCD}$. Similar to light flavors in our QCD, the dark quark masses have been assumed to be much lighter than $\Lambda_{\rm dQCD}$ and their contributions to the dark baryon mass can be neglected. We show the values of $m_D$ for different models in the last column of Table~\ref{tab:fixed-point-values}.

Before we present the numerical results of the scale of $M$ for different models, we first provide an approximate and analytic calculation. Below the scale $M$ and using only the one-loop beta function, 
%It is possible to obtain an analytic estimate for the ratio of DM masses, or more precisely, for the %ratio of confinement scales, using the RGE evolution at one loop, below the scale $M$. Then, 
the running of the couplings can be solved analytically and is given by
\begin{align}
	\alpha^{-1}_{i}(\mu) & \approx \alpha^{-1}_i(M) - \frac{\tilde{\beta}_i}{2\pi} \log \left( \frac{\mu}{M}\right)\,,
\end{align}
where $\tilde\beta_i = (\frac{2}{3} n_{f_i} -11)$. Using that the coupling at the confinement scale is $\alpha_i(\Lambda_i) = \pi/4$ and that $\alpha_i(M) = \alpha_i^*$, we can solve these equations for the confinement scales $\Lambda_i$ and obtain the ratio of the two confinement scales as
\begin{align}
	\frac{\Lambda_{\rm QCD}}{\Lambda_{\rm dQCD}} & \approx 
%	e^{\left(\frac{2 \pi}{\tilde\beta_c}\left(\frac{4}{\pi}-(\alpha_c^*)^{-1}\right)-\frac{2 \pi}{\tilde\beta_d}\left( \frac{4}{\pi} - (\alpha_d^*)^{-1}\right)\right)} \approx 
%	e^{\left(\frac{2 \pi}{\tilde\beta_d}(\alpha_d^*)^{-1}-\frac{2 \pi}{\tilde\beta_c}(\alpha_c^*)^{-1}\right)
%	e^{\left(\frac{2\pi}{\tilde{\beta}_c \,\alpha^*_c} - \frac{2\pi}{\tilde{\beta}_d \,\alpha^*_d} \right) }
	e^{\frac{2\pi}{\tilde{\beta}_c \,\alpha^*_c} \left(1 - \frac{\tilde{\beta}_c \,\alpha^*_c} {\tilde{\beta}_d \,\alpha^*_d}   \right)  }
	\,,
	\label{eq:confinement-ratio}
\end{align}
where we have used the perturbative IRFP gauge couplings with $\alpha_i^* \ll \pi/4$. Without a delicate cancellation between the two terms in the parenthesis, the ratio is ${\cal O} [e^{2\pi/(\tilde{\beta}_c \,\alpha^*_c})] \sim {\cal O} (10^{-4})$ for $\tilde{\beta}_c=-7$ and $\alpha^*_c \approx 0.1$. A mild cancellation around 20\% for the two terms in the parenthesis can have $\Lambda_{\rm dQCD}/\Lambda_{\rm QCD} \sim 5$ and comparable confinement scales.

% With $\beta_c = 7$ and $\alpha_c^* \gtrsim 10$, it follows that there can be sizeable differences in the scale unless some cancellation happens between the two terms in the exponent. More precisely, if we want $\Lambda_{\rm QCD}/\Lambda_{\rm dQCD} \lesssim 10$, we find the condition
%
%\begin{align}
%	\frac{\tilde\beta_c}{\tilde\beta_d} (\alpha_d^*)^{-1} - (\alpha_c^*)^{-1} \lesssim \frac{\tilde\beta_c}{2\pi} \log(10) \approx 2.6\,.
%\end{align}
%If $\alpha_d^* \geq \alpha_c^*$, the left hand side is of order $(\alpha_c^*)^{-1}$. For $M$ at the TeV scale, this is ${\cal O}(10)$, such that only a mild cancellation is required. We therefore expect that models with $\alpha_c^* \lesssim \alpha_d^* \lesssim 2 \alpha_c^*$ and $M\lesssim$~few~TeV will lead to a DM mass close to the proton mass scale. 

The ordinary QCD coupling running from $M_Z$ to the decoupling scale $M$ is determined by the known SM matter. Higher values of $M$ will lead to smaller values of $\alpha_c^*$ and require more tuning in the parenthesis of Eq.~(\ref{eq:confinement-ratio}), and are therefore disfavored. So, to have comparable $\Lambda_{\rm dQCD}$ and $\Lambda_{\rm QCD}$, the decoupling scale $M$ is preferred to be close to $M_Z$. 

%It also follows that models with larger $M$, corresponding to smaller values of $\alpha_c^*$, require more tuning, and are therefore disfavored. Similarly, models with $\alpha_d^* < \alpha_c^*$ require more tuning and are unlikely to give DM masses in the $1-10$~GeV range. 

\begin{figure}[th!]
\begin{center}
\hspace*{-0.75cm}
\includegraphics[width=0.6\textwidth]{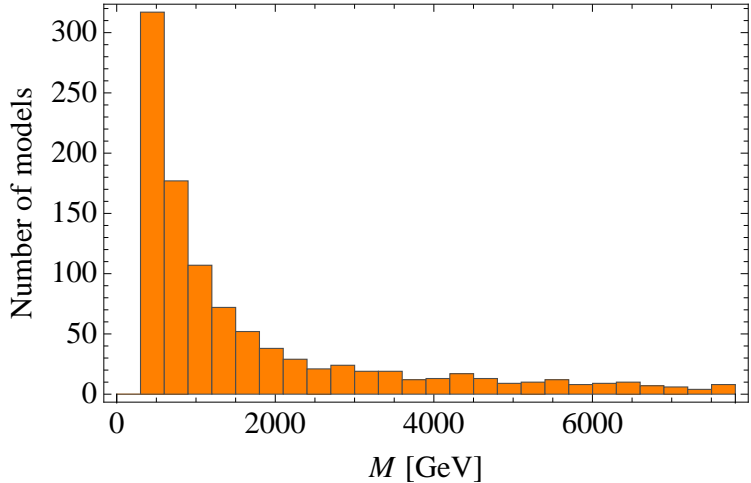}
\caption{The distribution of numbers of models in terms of the decoupling scale $M$, after satisfying the requirement of $1.5 < m_D/m_p < 15$. The lower limit of $M$ is related to requiring $\alpha_s^* \leq 0.1$.}
\label{fig:histogram}
\end{center}
\end{figure}

Given that the number of models that are asymptotically free in the far UV and exhibit IRFPs is finite, we can also analyze the distribution of DM masses numerically. Requiring $0.05 \leq \alpha_s^* \leq 0.1$ and a perturbative $\alpha_d^*$ leaves ${\cal O}(10^4)$ models for which we analyze the correlation between the scale $M$ and the dark baryon mass. If we imagine that $n_D/n_B = {\cal O}(1)$, a model is viable if $1.5 < m_D/m_p < 15$, such that the experimental value of $\Omega_{\rm DM}/\Omega_{\rm B}$ can be explained up to a range of a factor of three, leaving some room for numerical uncertainties. 
%
%There are around one million models that have both couplings to be asymptotic-free in UV and provide infrared fixed points to explain dark baryon masses. Requiring $0.05 \leq \alpha_s^* \leq 0.1$ and a perturbative $\alpha_d^*$, the number of models reduces to tens of thousand. To understand the correlation between $M$ and the dark baryon mass, we further require $1.5 < m_D/m_p < 15$, imagining that $n_D/n_B = {\cal O}(1)$, such that the experimental value of $\Omega_{\rm DM}/\Omega_{\rm B}$ can be explained up to a range of a factor of three. 
We show the distribution of numbers of models in $M$ in Fig.~\ref{fig:histogram}. 
As expected from our analytical estimates, in order to explain the experimental value of $\Omega_{\rm DM}/\Omega_{\rm B} \sim 5$, lower values of $M$ are clearly preferred. For the majority of models, $M \lesssim 2$~TeV, which is the mass scale of particles charged under both QCD and dark QCD and also determines the interaction strength between these two sectors.

%
%with the majority of models having $M \lesssim 2$~TeV. This mass scale determines the interaction strength between these two sectors as well as the prospects for producing the particles at collider experiments. 
%
%We see from Fig.~\ref{fig:histogram} that to explain the experimental value of $\Omega_{\rm DM}/\Omega_{\rm B} \sim 5$, the decoupling scale $M$ prefers to have a lower value. For the majority of models, $M \lesssim 2$~TeV, which is the mass scale of particles charged under both QCD and dark QCD and also determines the interaction strength between these two sectors.

%%%%%%%%%%%%%%%%%%%%%%%%%%%%%%%%%%%%
\section{Asymmetry from Leptogenesis}
\label{sec:leptogenesis}
%%%%%%%%%%%%%%%%%%%%%%%%%%%%%%%%%%%%
Having discussed the relation between the dark baryon and ordinary baryon masses, we now turn to the question of obtaining $n_D \sim n_B$. While there are many models to achieve this goal, we only present one simple renormalizable model following the leptogenesis idea~\cite{Fukugita:1986hr} and use it as a guidance for dark QCD phenomenologies. Leptogenesis is a well known mechanism to explain the baryon asymmetry of the Universe (BAU). It uses CP-violating, out-of-equilibrium decays of heavy right-handed neutrinos, $N_i$, to generate a lepton asymmetry at high scales. This lepton asymmetry is then partially transferred into asymmetry in the quark sector through electroweak sphaleron processes. 

In addition to the lepton asymmetry, it is also possible to generate an asymmetry of other quantum numbers from $N_i$ decays~\cite{An:2009vq,Falkowski:2011xh}. In the following we show a model to generate both the BAU and the dark BAU at the same time. Differently from Ref.~\cite{An:2009vq,Falkowski:2011xh}, our model will have the baryon and the dark baryon asymmetries controlled by the same model parameters, and $n_D/n_B={\cal O}(1)$ can be achieved naturally.

The main idea to generate an asymmetry for a particle that can decay into ordinary baryons and dark baryons, so $n_B$ and $n_D$ can share the same source of asymmetry. The particles bi-fundamental of QCD and dark QCD are natural candidates for this. For instance, one can induce an asymmetry in a $(\overline{3}, 3)_{1/3}$ fermion $Y_1$, such that $\Delta n_{Y_1} \equiv n_{Y_1} - n_{\bar{Y_1}} \neq 0$. Note that we only write down the quantum numbers under $SU(3)_{\rm QCD} \times SU(3)_{\rm dQCD} \times U(1)_Y$, since all fields involved will be $SU(2)_{\rm weak}$ singlets. Since $Y_1$ carries both QCD and dark QCD colors, its decays will distribute the asymmetry evenly between the visible and the dark sectors. To generate the asymmetry via leptogenesis, we introduce a $(\overline{3},3)_{1/3}$ scalar $\Phi$ with Yukawa couplings:
\beqa
	{\cal L} \supset k_i \bar{Y}_1 \Phi N_i + {\rm h.c.}
	\label{eq:lag-Ni}
\eeqa
Here, $N_i$, $(i=1,2,3)$ are three heavy right-handed neutrinos with Majorana masses $M_i$ ($M_i<M_j$ for $i<j$) that could  also be responsible for generating small SM neutrino masses through the seesaw mechanism. Out of equilibrium decays of $N_1$ in the early Universe can generate asymmetries $\Delta n_{Y_1}\equiv n_{Y_1} - n_{\bar{Y}_1}$ and $\Delta n_\Phi = -\Delta n_{Y_1}$, provided that $\mathfrak{Im}[k_1^2 (k_2^*)^2]$ is nonzero. An estimate of the amount of asymmetry generated from these decays will be presented later. 

Additional fields and couplings are required to allow the asymmetry to be transferred to baryons and dark baryons. We introduce a second bitriplet fermion $Y_2$ transforming as $(\overline{3},3)_{-2/3}$, and Yukawa couplings
\beqa
	{\cal L} \supset  \kappa_{1} \, \Phi\, \bar{Y}_1^c\, Y_2  + \kappa_{2}\, \Phi \,\bar{Y}_2 \, e_R  + \kappa_3\, \Phi \, \bar{X}_L \, d_R  + {\rm h.c.} \, ,
       \label{eq:lag-Yi}
\eeqa
where $Y_1^c = {\cal C}\, Y_1^T$ and ${\cal C}$ is the charge conjugation operator. Here, $e_R$ and $d_R$ are the right-handed SM charged leptons and down-up quarks, respectively, with the flavor indices suppressed. For simplicity, we assume that the $\Phi$ field is lighter than $Y_i$, but with a small mass hierarchy. Then, we have the decay chains $Y_1 \to \bar{Y}_2+\Phi^\dagger$ followed by $Y_2 \to \Phi+ e_R$ and $\Phi \to X_L+ \bar{d}_R$. The asymmetries that are initially stored in the $\Phi$ and $Y_1$ fields are distributed as follows:
\begin{align}
	\Delta n_{d_R} &\equiv 3\,  n_{B} = 3\, \Delta n_{Y_1}\,, \\
	\Delta n_{e_R} &\equiv n_L = - \Delta n_{Y_1}\,, \\
	\Delta n_{X} & \equiv 3\, n_{D} = -3\, \Delta n_{Y_1}\,,
\end{align}
where we have taken into account that each (dark) quark carries $1/3$ of the (dark) baryon number. The $B-L$ asymmetry is given by $ n_{B}- n_L = 2 \Delta n_{Y_1}$. Weak interaction, Yukawa interactions and electroweak sphaleron processes will redistribute the asymmetries across SM quarks and leptons. Assuming that the lepton flavors equilibrate, we use the well-known relation $ n_{B} = 28/79 \,n_{B-L}$~\cite{Harvey:1990qw,Chung:2008gv} to obtain the ratio of $n_D/n_B$ as 
\begin{align}
	\frac{|n_D|}{n_B} = \frac{79}{56}  \,.
	%\approx \frac{7}{5}\,. 
	\label{eqn:ratioapprox}
\end{align}
The interactions introduced in Eqs.~(\ref{eq:lag-Ni}) and (\ref{eq:lag-Yi}) conserve a dark matter ${\cal Z}_2$ symmetry. Under this ${\cal Z}_2$, we find the fields $X$, $\Phi$, $e_R$, $Y_1$ to be odd and $Y_2$ and $N_i$ to be even. So, the dark baryon constructed from three $X$ fields is ${\cal Z}_2$ odd and stable.

Before we calculate the energy density ratio, we digress into discussing how to obtain $\Delta n_{Y_1}$ from leptogenesis. The lightest right-handed neutrino, $N_1$, must decay sufficiently out of equilibrium. This is possible if the decay rate $\Gamma_{N_1} = 9 |k_1|^2 M_1/(16 \pi)$ is not too different from the Hubble expansion rate of the Universe $H(T=M_1)$ at a temperature $T=M_1$.
%
%\begin{align}
%	\Gamma_{N_1} = \frac{9 |k_1|^2 M_1}{16 \pi}\,, \qquad H(T) = \frac{2}{3} \sqrt{\frac{g_\star \pi^3}{5} } \frac{T^2}{M_{\rm pl}} \,,
%\end{align}
%
%where $M_{\rm pl} = 1.22\times 10^{19}$~GeV is the Planck mass, and $g_\star$ is total number of degrees of freedom in the universe. Depending on the exact particle content we have $g_\star \approx 300$, compared to $g_{\star,\rm SM}= 106.75$.
This condition $\Gamma_{N_1} \sim H(M_1)$ roughly translates to $|k_1|^2 \sim M_1/(10^{17}$~GeV), so it can be easily satisfied for a $N_1$ mass below the Planck scale. 

The CP-asymmetry in the decay $N_1 \to Y_1 \Phi^\dagger$ can be inferred from the known leptogenesis result~\cite{Covi:1996wh}. In the hierarchical limit, $M_2 \gg M_1$, and neglecting finite temperature corrections, it is given by~\cite{Beneke:2010wd}
\begin{align}
	\epsilon &= \frac{\Gamma(N_1 \to Y_1 \Phi^\dagger) - \Gamma(N_1 \to \bar{Y}_1 \Phi)}{\Gamma(N_1 \to Y_1 \Phi^\dagger) + \Gamma(N_1 \to \bar{Y}_1 \Phi)}
	\approx - \frac{3}{2}\frac{1}{8 \pi} \frac{{\mathfrak Im} [k_1^2 (k_2^*)^2] }{|k_1|^2} \frac{M_1}{M_2}\,.
\end{align}
In the strong washout regime, $\Gamma_{N_1} \gg H(M_1)$, the final asymmetry can be estimated as~\cite{Buchmuller:2004nz,Fong:2013wr}
\begin{align}
	Q_{Y_1} (\infty) &= \frac{\pi^2}{6 z_f K_1} \epsilon \,Q_{N_1}^{\rm eq}(0)\,,
\end{align}
where $Q_i = n_i/s$ are the entropy normalized particle densities, $K_1 = \Gamma_{N_1}/H(M_1)$ and $z_f$ is the freeze-out temperature where the washout decouples, with $z_f \sim 7-10$ for $K_1 = 10-100$. The equilibrium $N_1$ density at high temperatures is approximately given by $Q_{N_1}^{\rm eq}(0) \approx 4/g_\star$, with $g_\star \approx 300$ in our model. Choosing $M_1 = 10^{13}$~GeV, $|k_1| = |k_2| = 0.1$, and $M_2 = 10 M_1$, we have $Q_{Y_1}(\infty) \approx 2 \times 10^{-9}\sin(2\varphi)$, where $\varphi$ is the relative CP phase in the couplings $k_{1,2}$.
In comparison, the observed baryon to entropy ratio today is $9 \times 10^{-11}$~\cite{Fong:2013wr}. Therefore it is easy to see that a large enough asymmetry can be generated to explain the observed baryon and dark baryon asymmetries of the Universe.  

\begin{figure}[th!]
\begin{center}
\hspace*{-0.75cm}
\includegraphics[width=0.6\textwidth]{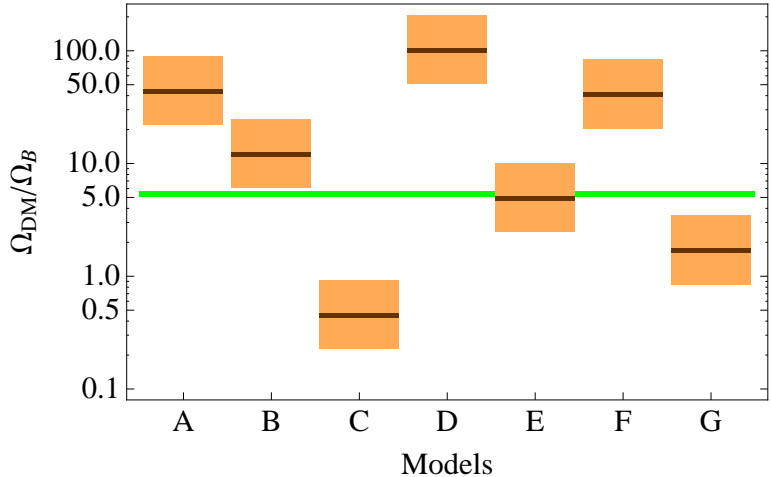}
\caption{The ratios of the dark baryon energy density over the ordinary baryon energy density for different models in Table~\ref{tab:fixed-point-values}. The dark lines are the ratios $\Omega_{\rm DM}/\Omega_{\rm B}$ calculated using Eq.~(\ref{eqn:DMratio}) for different models, while the orange (grey) bands are obtained by letting the dark baryon mass vary between $1/2$ and $2$ times the estimated value, to account for the uncertainty of the nonperturbative estimation of $\Lambda_{\rm dQCD}$ (a more precise calculation could be done at Lattice~\cite{Appelquist:2013ms}). The green line is the measured value of $\Omega_{\rm DM}/\Omega_{\rm B}$ from the Planck Collaboration. 
}
\label{fig:DM-energy-density}
\end{center}
\end{figure}

After discussing asymmetry generation, we now come back to calculate $\Omega_{\rm DM}/\Omega_{\rm B}$, which is simply given by
\beqa
	\frac{\Omega_{\rm DM}}{\Omega_{\rm B}} = \frac{n_D\,m_D}{n_B\, m_p} \approx \frac{79}{56} \frac{m_D}{m_p}\,.
	\label{eqn:DMratio}
\eeqa
Assuming the same $n_D/n_B=79/56$ for all models, we show the dark matter energy densities in Fig.~\ref{fig:DM-energy-density} for the representative models in Table~\ref{tab:fixed-point-values}. Note that while we show a variety of models here, only models D, E, F and G have the necessary particle content to implement the asymmetry mechanism in this section. 
Among different models, the model ``$E$" has a dark matter mass around 3.5~GeV and the ratio $\Omega_{\rm DM}/\Omega_{\rm B} \approx 4.9$, which is very close to the measured value from the Planck Collaboration.

A prominent issue in asymmetric dark matter model building is that the dark matter - antidark matter annihilation rate must be sufficiently efficient to prevent a large symmetric relic density. In our model, this potential problem is naturally solved because the dark baryon and antibaryon annihilation into dark pions is very efficient,  similar to the proton and antiproton annihilation in the SM. The dark pions do not carry dark baryon number, so they can decay into SM particles (unless they have their discrete symmetries for stability, for instance in~\cite{Kilic:2009mi, Bai:2010qg, Buckley:2012ky}, which we do not consider here). We discuss their properties in the next section for the phenomenology of our model.

%--------------------------------------------------------------------%
\section{LHC and dark matter phenomenology} 
\label{sec:pheno}
%--------------------------------------------------------------------%
So far, the chiral symmetry, $SU(n_{f_d})_L \times SU(n_{f_d})_R$,  associated with the dark quarks is unbroken. To provide masses to the otherwise massless Nambu-Goldstone bosons or dark pions, $\pi_d$, we adopt the Higgs portal and introduce the dark-flavor-blind interactions,  $\bar{X} X H^\dagger H/\Lambda$, which can be easily UV-completed by introducing a gauge singlet field $S$ with two couplings $\bar{X}X S$ and $S H^\dagger H$. The dark pion mass has the approximate relation $m^2_{\pi_d} f_{\pi_d}^2 \sim m_{X} \Lambda_{\rm dQCD}^3$, with the dark quark mass $m_X \sim v^2_{\rm EW}/\Lambda$.~\footnote{The Yukawa coupling of dark quarks to the Higgs boson is $\sim m^2_{\pi_d} f_{\pi_d}^2/(v_{\rm EW}\,\Lambda_{\rm dQCD}^3)$, which is suppressed by at least a power of $f_{\pi_d}^2/\Lambda_{\rm dQCD}^2 \sim 1/(4 \pi)^2$ and will not affect the SM Higgs properties in a significant way.} The dark pion masses are controlled by additional UV parameters and can well be below the dark baryon mass.~\footnote{Other phenomenological studies of the dark pion or a general dark QCD sector can be found in~\cite{delaMacorra:2002tk,Hur:2007uz,Higaki:2013vuv}.}

The dark QCD and our QCD sectors are coupled to each other through the bifundamental particles, whose mass scale $M$ is crucial for the phenomenology of this class of dark QCD models. Integrating out the bifundamental $\Phi$ field, one can generate the operator $\kappa_3^2 \,X_L \overline{d}_R\,d_R \overline{X}_L/M_{\Phi}^2$. After Fiertz transformation, this operator becomes $\kappa_3^2\,\overline{X}_L \gamma_\mu X_L\, \overline{d}_R \gamma^\mu d_R/M_{\Phi}^2$. First of all, one can see that the dark parity is broken and the dark pion can decay into SM quarks from the operator, $i \kappa_3^2\, f_{\pi_d}\,m_d\,\pi_d \,\overline{d} \gamma_5 d/M_{\Phi}^2$, using the dark chiral Lagrangian. For $\Lambda_{\rm dQCD} > \Lambda_{\rm QCD}$, the decay width of $\pi_d$ is estimated to be $\kappa_3^4\,f_{\pi_d}^2 m_d^2 m_{\pi_d}/(32\pi M_\Phi^4)$. For $M_\Phi/\kappa_3 \sim 1$~TeV, the dark pion is generically a stable particle at colliders unless $\pi_d$ is heavy enough to decay into strange quarks.
%
%\begin{align}
%	\Gamma(\pi_d \to d \bar{d} ) & = \frac{1}{16\pi} \frac{f_{\pi_d}^2 m_d^2 }{2 M_\Phi^4} m_{\pi_d} \left(1 - \frac{2 m_d^2}{m_{\pi_d}^2}\right) \sqrt{
%1- \frac{4 m_d^2 }{ m_{\pi_d}^2} } .
%\end{align}
%
When the dark pion mass is below $3\,m_\pi$, it can only decay into a pair of photons at loop level or high-multiplicity final state via off-shell pions and has an even longer lifetime. 

The effective operator, $\kappa_3^2\,\overline{X}_L \gamma_\mu X_L\, \overline{d}_R \gamma^\mu d_R/M_{\Phi}^2$, can also be used to induce both dark matter-nucleon spin-independent and spin-dependent scattering. For the dominant spin-independent scattering, the matrix element for scattering off a proton or neutron is given by~\cite{Goodman:1984dc} ${\cal M}_{p,n} = \kappa_3^2/(4M_\Phi^2) J^0_X J^0_{p,n}$,  where $J^0_X = \langle D | \overline{X} \gamma^0 X | D \rangle \approx 3$ and $J^0_{p,n} = \langle p,n| \overline{d} \gamma^0 d|p,n\rangle \approx 1,2$. Then the spin-independent  dark baryon-neutron cross section is calculated to be
\beqa
\sigma^{\rm SI}_{D-n} = \frac{2^2\,3^2\,\kappa_3^4\, \mu^2_{D-n}}{16\,\pi\,M_\Phi^4} =  \left( \frac{1~\mbox{TeV}}{M_\Phi/\kappa_3} \right)^4 \times 3\times 10^{-40}~\mbox{cm}^2 \,,
\eeqa
where $\mu_{D-n}$ is the reduced mass of the dark baryon and ordinary neutron system. For $m_D\approx 3.5$~GeV and $M_\Phi/\kappa_3= 1$~TeV, the cross section is close but below the current limits from light dark matter searches~\cite{Angle:2011th,CDEX1306}. 

In our model, we have additional particles charged under the SM QCD with masses at the decoupling scale $M$. The lightest additional QCD charged state $\Phi$ can be produced in pairs at the LHC. Each $\Phi$ can decay into one quark and one dark quark, $\Phi \rightarrow X \bar{d}_R$. After hadronization, the ordinary quark will behave as a jet at colliders. The story for the dark quark is slightly different. After hadronization in the dark sector, both dark baryons and dark mesons can exist in the final state. If dark pions are stable particles at colliders, the total dark jet behaves as missing energy. The final signal is two QCD jets plus missing transverse energy, well covered by the current SUSY search~\cite{ATLAS:2012ona,Chatrchyan:2013lya}. Recasting the results in Ref.~\cite{Chatrchyan:2013lya} with 11.7 fb$^{-1}$ at 8 TeV and including the multiplicity factor for the $\Phi$ field with respect to the squark production in SUSY models, the current constraint is $M_{\Phi} \gtrsim 600$~GeV.

On the other hand, if the dark pions decay into SM quarks inside detectors, only a fraction of the dark jet momentum contributes to the transverse missing energy momentum and a dedicated search beyond the SUSY search is required. The closest signature to search for the $\Phi$ is the four-jet final state and paired dijet resonance search. Compared to the limits of stop in Fig.~3 of Ref.~\cite{Chatrchyan:2013izb}, the constraint on the $\Phi$ mass is $M_{\Phi} \gtrsim 400$~GeV after taking into account of the multiplicity factor. The actual constraints should be weaker because not all $X$-dark-jet energy is registered in the calorimeter. 

%--------------------------------------------------------------------%
\section{Conclusions}
\label{sec:conclusion}
%--------------------------------------------------------------------%
To conclude, we have provided a simple explanation for why the dark QCD scale could be related to the SM QCD scale, or why the dark baryon mass could be similar to the proton or neutron mass. IRFP points from gauge coupling runnings play an essential role. The two gauge couplings are related to each other until a decoupling scale $M$, where particles charged under both gauge groups get their masses. We have scanned all the models with fundamental representations and asymptotic-free runnings in the far UV, and found that the majority of models with $m_D\sim m_p$ have a decoupling scale $M$ around one TeV. We have also provided a simple renormalizable model to explain $n_D \sim n_B$ based on the leptogenesis mechanism and a sharing of dark baryon and ordinary baryon asymmetries. Therefore, a dark QCD with fixed-point gauge couplings in the infrared provides a dynamical explanation for the similarity of dark matter and ordinary matter energy densities that is observed in nature.

\subsection*{Acknowledgments} 
We would like to thank Randy Cotta and Takemichi Okui for useful discussions and comments. 
Y. Bai is supported by startup funds from the UW-Madison. P.~Schwaller is supported in part by the U.S. Department of Energy, Division of High Energy Physics, under grant numbers DE-AC02-06CH11357 and DE-FG02-12ER41811. We also would like to thank the Kvali Institute for Theoretical Physics, U. C. Santa Barbara, where part of this work was done. This research was also supported in part by the National Science Foundation under Grant No. NSF PHY11-25915.

%%%%%%%%%%%%%%%%%%%%%%%%%%%%%%%%%%%
\bibliography{InfraredDM}
\bibliographystyle{JHEP}

 \end{document}